\documentclass{ws-ijmpd}
\RequirePackage{graphicx}
\usepackage{epstopdf}
\usepackage{bm}
\usepackage{amsfonts}

\begin{document}
	
	\title{\bf  Modified Power law Inflation: solution to the graceful exit problem and improvement of dark energy models}
	\author{Prasenjit Paul}
	\address{Department of Physics, Government College of Engineering and Ceramic Technology, Kolkata 700 010, West Bengal, India \\Department of Physics, Indian Institute of Engineering Science and Technology, Howrah \\711 103, West Bengal, India \\prasenjit071083@gmail.com}
	
	\author{Rikpratik   Sengupta}
	\address{Department of Physics, Government College of Engineering and Ceramic Technology, Kolkata 700 010, West Bengal, India \\ rikpratik.sengupta@gmail.com}
	
	\author{Saibal Ray}
	\address{Department of Physics, Government College of Engineering and Ceramic Technology, Kolkata 700 010, West Bengal, India \\ saibal@associates.iucaa.in}
	
	\maketitle
	
	\begin{abstract}
		We study power law inflation (PLI) with a monomial potential and find a novel exact solution. It is well known that conventional PLI with exponential potential is inconsistent with the Planck data. Unlike the standard PLI, present model does not suffer from graceful exit problem and it agrees fairly well with recent observations. We have calculated the spectral index and the tensor-to-scalar ratio which are in very good agreement with recent observational data and also comparable with other modified inflationary models. A technique has been used which shows that the large cosmological constant reduces with expansion of the Universe in case of the power law inflation. The coupling of the inflaton with gravitation is the main point in this technique. The basic assumption here is that the two metric tensors in the gravitational and the inflaton parts correspond to different conformal frames which is in contradiction with the conventional power law inflation where the inflaton directly coupled with the background metric tensor. This fact has direct application to different dark energy models and assisted quintessence theory.
	\end{abstract}
	

	\section{Introduction}
	Standard cosmology has some unanswered problems, such as flatness, horizon problems etc. It is possible to address solution to these problems if one consider that there exist a phase of inflation at early Universe. Over the years numerous models have been proposed which generate such an inflationary phase. The basis of the old models~\cite{guth,al,linde1,linde2} were the Universe begins in a thermal equilibrium state and there exist a Higgs field which is the inflaton, procured through spontaneous symmetry breaking, a huge amount of energy density for a properly selected effective potential. In case of the theories regarding inflation Linde introduced a new paradigm, viz. the `chaotic inflation'~\cite{linde3,linde4}. In single-field inflationary models it has been assumed that there exist sufficiently flat potential of the scalar field which dominates the energy density of the Universe. On the other hand, if more than one scalar field are involved, inflation is possible even when the potential is not flat as discussed elaborately in case of hybrid inflation~\cite{linde5,bellido} and assisted inflationary models~\cite{linde6}. So in these cases no {\it ad hoc} assumptions are required on the shape of the potential. 
	
	However, there are still different unsolved problems in inflationary cosmological scenario. Among those problems the {\it Achilles heel} is probably the cosmological constant problem as there exist a huge gap between observational value of the vacuum energy density  and its theoretical prediction~\cite{weinberg,carroll,padmanabhan}. The upper bound of the observational data~\cite{komatsu1,komatsu2} of vacuum energy density ($\rho_{\Lambda}$) are 120 order smaller than theoretically predicted values for different models. In the work of Weinberg~\cite{weinberg} different approaches were stated for the solution to this problem. A model~\cite{wetterich} has been proposed where energy momentum tensor adjusted dynamically such that cosmological constant become time-dependent. The large vacuum energy density is necessary in different inflationary models to produce enough inflation. It can be predicted that the cosmological constant ($\Lambda$) may have a large value at early times, consistent with the theoretical predictions\cite{Paul}. Few phenomenological $\Lambda$ models focus on the importance of variable time-dependent $\Lambda$~\cite{saibal1,saibal2}. Some definite dynamical models of $\Lambda$ have been selected for inspecting the nature of dark energy~\cite{saibal3,saibal4}. The fact that the variability of cosmological constant in the form of dark energy has been discussed in~\cite{bharat}. A simple toy model has been provided which keeps room for the end of inflation to avoid the possible deadlock~\cite{prasenjit} and the results agree well with the observational data~\cite{wmap,planck1,planck2,keck}. There is a recent work which shows dynamical process decaying vacuum energy density from inflation to a radiation phase followed by dark matter and vacuum regimes~\cite{basilakos}. In connection to inflationary cosmology some interesting works with dark energy are available in the literature~\cite{DK1998,DK2000,DK2001}.
	
	In general relativity uses of conformal re-scalings and conformal techniques have been done for a long time. Different models of inflation based on non-minimal gravity can be treated in a similar manner as standard inflationary analyses by means of conformal transformations~\cite{kalara}. There are application of conformal transformation to a general class of single field inflation models to gravity and non-standard kinetic terms with non-minimal coupling~\cite{kubota}. Over the years technique  of Weyl or conformal transformation has been used by which cosmological constant reduces as the time progresses~\cite{zha,guen,faraoni,darabi1}. 
	
	In the present work we mainly target on the duration of inflationary phase and use a progressive technique~\cite{darabi2,bis1}. This progressive technique is important in the quintessence theory as it predict that dark energy is a new force and will finally fade away just as it arose. We take two assumptions in our model. The first assumption is that there exists a large effective cosmological constant at early Universe. In the present work we model the cosmological constant term by an inflaton which is a minimally coupled scalar field and the corresponding potential can acquire contributions from different fields characterizing particle physics. The second assumption deals with the fact that there exist a conformal background metric which couples with the large energy density, i.e. it is related to gravitational coupling of the cosmological term. For the presence of the gravitational coupling, the conformal factor comes out to be a damping factor in the cosmological term which is responsible for the reduction of the latter during inflation. As the potential associated with the scalar field reduces during inflationary phase, the Universe undergoes a power law inflation (PLI) not the de Sitter inflation. In the present work, PLI justifies small value of the cosmological constant (i.e. the vacuum energy) at late times as well as solves the horizon and the flatness problems. 
	
	In Sect. 2 basic features of slow roll parameters in conventional PLI with exponential potential are stated. In Sect. 3, the progressive technique using the conformal transformation for a monomial potential has been discussed. We give the field equations and evaluate their exact solutions. It has been observed that the solutions exhibit a PLI for a power-law potential of the scalar inflaton. Sect. 4 deals with the comparison of our result in this model with recent observational data. In Sect. 5, we provide some concluding remarks.

	\section{Slow-roll parameters in the conventional PLI}
	In a single-field inflationary models the Lagrangian density can be written as
	\begin{equation}{\mathcal{L}}(g_{\alpha\beta},
	\psi)=\frac{1}{2}g^{\alpha\beta}\nabla_{\alpha}\psi\nabla_{\beta}\psi+V(\psi),
	\label{0-2}
	\end{equation}
	where $\psi$ is the inflaton field and the potential function is represented by $V(\psi)$. 
	
	The full action in the unit of, $\hbar=c=1$ is given by
	\begin{equation}
	S=\frac{1}{16\pi G} \int d^{4}x \sqrt{-g} {\mathcal{R}} - \int d^{4}x \sqrt{-g}\mathcal{L}\label{0-1}
	\end{equation}
	
	In above equation the first term represents the Einstein-Hilbert action.
	
	If we consider the Universe to be isotropic and homogeneous then the evolution of the inflaton field $\psi(t)$ and the scale factor $a(t)$ can be written by the Friedmann equation
	\begin{equation}
	3H^2=k\left[\frac{1}{2}\dot{\psi}^2+V(\psi)\right],
	\label{0-3}
	\end{equation}
	
	\begin{equation}
	\ddot{\psi}+3H\dot{\psi}=-V'(\psi), \label{0-4}
	\end{equation} 
	where Hubble parameter $H=\frac{\dot{a}}{a}$,~$k=8\pi G$. The dot represent differentiation with respect to $t$ whereas the prime indicate differentiation with respect to $\psi$. When the slow roll parameters are small we will have a period of accelerated expansion, i.e. inflation and the corresponding slow-roll parameters can be written as~\cite{slow}
	\begin{equation}
	\epsilon(\psi)=\frac{m_p^2}{2}\left[\frac{V'(\psi)}{V(\psi)}\right]^2,
	\label{0-5}
	\end{equation}
	
	\begin{equation}
	\eta(\psi)=m_p^2\left[\frac{V''(\psi)}{V(\psi)}\right].
	\label{0-6}
	\end{equation} 
	
	Here  ${m_p}^2=\frac{1}{G}$, $m_p$ is the Planck mass. As the slow-roll parameters are small it is justified to ignore the time-derivatives of $\psi$ in Eqs. (\ref{0-3}) and (\ref{0-4}) and the potential term is much greater than the kinetic contribution, i.e.
	$\frac{1}{2}\dot{\psi}^2<<V(\psi)$. So, Eq.  (\ref{0-3}) becomes $3H^2\approx k V(\psi)$. Therefore when $\psi \approx$ constant, the energy density corresponding to the scalar field has a constant value which gives a de sitter solution. 
	
	Beside the exponential acceleration, non-exponential accelerated expansion also can perform the work ~\cite{abbott,lucchin,sahni}. One may consider, e.g. power law expansion and the corresponding scale factor is given by, $a(t)\sim t^q$ where the power law index $q$ is larger than unity ($q>1$). In conventional PLI for a canonical scalar field has an exponential potential of the form
	$V(\psi)=V_0e^{-\sqrt{\frac{2}{q}}(\frac{\psi}{M_p})}$ with $V_0$ as a constant~\cite{halliwell,barrow,yokoyama,burd,liddle}. With the exponential potential slow roll parameters from equations (\ref{0-5}) and (\ref{0-6}) can be written as
	\begin{equation}
	\epsilon=\frac{1}{q}, \label{0-7}
	\end{equation}
	
	\begin{equation}
	\eta=\frac{2}{q}. \label{0-8}
	\end{equation} 
	
	During slow roll paradigm $\epsilon, \eta~<<1$ hence $q>>1$. In the case of de Sitter inflationary paradigm, exponential expansion terminates when the slow-roll approximation is no longer applicable. Since in PLI, $q$ is a constant parameter hence termination of inflation is not clear. So an exit mechanism should be incorporated to the whole inflationary 
	scenario such that as expansion proceeds and the inflationary phase matches with the standard hot Big Bang model, i.e. the radiation or the matter dominated region.
	
	Any inflationary model is usually characterized by the scalar spectral index $n_s$ and the tensor-to-scalar ratio $r$~\cite{Starobinsky1,Starobinsky2}. These two parameters are related to the slow-roll parameters ($\epsilon, \eta$) as
	\begin{equation}
	n_s-1 = 2\eta-6\epsilon, \label{0-9}
	\end{equation}
	
	\begin{equation}
	r = 16\epsilon. \label{0-10}
	\end{equation} 
	
	In case of PLI, we have
	\begin{equation}
	n_s-1 = -\frac{2}{q}, \label{0-11}
	\end{equation}
	
	\begin{equation}
	r = \frac{16}{q}. \label{0-12}
	\end{equation}
	
	The Planck data~\cite{planck1} combining with the Wilkinson Microwave Anisotropy Probe (WMAP) result~\cite{wmap} require that the value of scalar spectral index ($n_s$) be in the range $n_s\in [0.945, 0.98]$ and the tensor-to-scalar ratio, $r<0.11$ (In the model Planck TT + lowP + BAO). Constraint from a combination of Planck, BICEP2 and	Keck Array data is $r<0.07$ \cite{planck2,keck}. The observational limits on $n_s$ is equivalent to the limits of the power law index as $38\leq q \leq 101$ and which corresponds to  $0.16 < r < 0.43$. This range of $r$ lies beyond the above-mentioned ranges. In Fig. 2 the outermost and middle contour represents Planck+WP data and the innermost contour corresponds to Planck+BICEP2+Keck Array data. The solid line which represents the PLI even lies completely outside the outer contour.
	
	So the conventional PLI has two drawbacks. First the graceful exit problem and second is the mismatch of the observations with the theoretical predictions. Hence, one require to clarify Lagrangian density corresponding to inflaton such that $n_s$, $r$ become consistent with the observational data~\cite{sahni,sahni1}.

	\section{Progressive technique: used as a tool of conformal transformation}
	Usually conformal transformation has been used as a mathematical tool for mapping of the equations of motion between a physical systems and a mathematically equivalent sets of equations making them easier to solve and computationally more convenient to study. In this model we take two assumptions regarding the gravitational coupling of the matter systems. Firstly, different types of matter couple with a particular metric. That means all types of field in different standard model couple in the similar manner to gravity irrespective of their huge variation in physical properties. Secondly, we have a unique metric, describing the background geometry. Thus the gravitational coupling is universal and equivalence principle support this fact having many observable results~\cite{dam1,dam2}. It has been also verified empirically many times since seventeenth century~\cite{will1,will2}. In this case the results has been extrapolated to total age of the Universe because all equivalence principle tests occur in a limited time interval (four hundred years since the time of Galileo). But there is a possibility that equivalence principle has been violated in some portion of the evolution of the Universe. On the other hand, all EP tests are also restricted in the Solar System. It is a well-known fact that there exist some screening techniques by which an anomalous gravitational coupling of matters can be obscured from experiments, e.g. if chameleon scalar field interacting with the matters~\cite{khoury1,khoury2}, then such an interaction can not be detected empirically. In this case, the local gravity constraints are suppressed in laboratory as the chameleon field is heavy. Meanwhile, it can be light enough in the low-density cosmological environment to have observable effects at the large scale.
	
	In this paper we try to consider a gravitational coupling to cool off the aforementioned assumptions. We also consider the fact that the two metrics for the matter and the gravitational sectors pertain to different conformal frames. In this case we take a minimally coupled scalar field as depicted by the Lagrangian density~(\ref{0-2}). As we consider contributions of the various fields of elementary particles, the potential of the inflaton $\psi$ have large effective masses which correspond to a large effective cosmological constant. Now, we consider in the full action the inflaton part pertain to a different conformal frame, specified by~\cite{hawking,wald}
	\begin{equation}
	\bar{g}_{\alpha\beta} =e^{-2\xi}  g_{\alpha\beta}, \label{1-1}
	\end{equation}
	
	\begin{equation}
	\bar{\psi} = e^{\xi} \psi. \label{1-2}
	\end{equation}
	
	From Eq. (\ref{1-1}) the inverse metric, $g^{\alpha\beta}$ and the determinant, $g$=det[$g_{\alpha\beta}$] transform as
	\begin{equation}
	\bar{g}^{\alpha\beta} =e^{2\xi}  g^{\alpha\beta},
	\end{equation}
	
	\begin{equation}
	\sqrt{-\bar{g}} = e^{-4\xi} \sqrt{-g}. \label{inverse}
	\end{equation}  
	
	We consider the conformal transformations as a local unit transformations~\cite{weyl,Dicke,Narlikar,reines,Hoyle,bek} with a space-time dependent conversion factor.  Usually, $\xi$ depends on space-time and also it is a smooth, dimensionless function, however later in the calculation we take that $\xi$ is a function of time only.  Hence the Lagrangian density corresponding to the inflaton can be written as
	\begin{equation}
	{\mathcal{L}}(\bar{g}_{\alpha\beta},
	\bar{\psi})=\frac{1}{2}\bar{g}^{\alpha\beta}\nabla_{\alpha}\bar{\psi}\nabla_{\beta}\bar{\psi}+V(\bar{\psi}).
	\label{1-3}
	\end{equation} 
	
	Using the fact that the Lagrangian is invariant under conformal transformation the full action from Eq. (\ref{0-1}) is given by
	\begin{eqnarray}
	S=\frac{1}{2k} \int d^{4}x \sqrt{-g} {\mathcal{R}} -\int d^{4}x
	\sqrt{-\bar{g}} {\mathcal{L}}(\bar{g}_{\alpha\beta} \bar{\psi}).
	\label{1-4}
	\end{eqnarray}
	
	In terms of $g_{\alpha\beta}$ and $\psi$ the action~(\ref{1-4}) becomes
	\begin{eqnarray}
	S =\frac{1}{2} \int d^{4}x \sqrt{-g}&\bigg\{&\frac{1}{k}{\mathcal{R}}-g^{\alpha\beta}
	\nabla_{\alpha} \psi \nabla_{\beta}\psi  
	-2\psi g^{\alpha\beta} \nabla_{\alpha} \psi
	\nabla_{\beta}\xi -\psi^2 g^{\alpha\beta} \nabla_{\alpha}
	\xi \nabla_{\beta}\xi\nonumber\\&& -V({e^{\xi}}\psi)e^{-4\xi}\bigg\}.
	\label{1-6}
	\end{eqnarray} 
	
	The action functional thus obtained depends on two scalar fields, viz.~$\xi$,~$\psi$ which are dynamical in nature with a term~\cite{k1,k2} of mixed kinetic type. This type of system has important application in the formulation of assisted quintessence~\cite{kim,ohashi,amendola,barros} and also to ameliorate different dark energy models~\cite{k2,chimento,barros2}.  
	
	We may write
	\begin{equation}
	\psi g^{\alpha\beta} \nabla_{\alpha}\psi \nabla_{\beta}\xi = \nabla_{\alpha} (\psi \xi g^{\alpha\beta}\nabla_{\beta}\psi)-\xi g^{\alpha\beta} \nabla_{\alpha}
	\psi\nabla_{\beta}\psi-\psi \xi \Box\psi. \label{surface}
	\end{equation}
	
	Using Eq. (\ref{surface}), the action~(\ref{1-6}) can be written as
	\begin{eqnarray}
	S =\frac{1}{2} \int d^{4}x \sqrt{-g} &\bigg\{&\frac{1}{k}{\mathcal{R}}-(1-2\xi)g^{\alpha\beta}
	\nabla_{\alpha} \psi \nabla_{\beta}\psi 
	- 2\psi \xi \Box \psi
	-\psi^2 g^{\alpha\beta} \nabla_{\alpha}
	\xi \nabla_{\beta}\xi \nonumber\\&&-V(e^{\xi}\psi)e^{-4\xi}\bigg\}.
	\label{1-6a} 
	\end{eqnarray} 
	
	When the slow-roll condition is valid we can write
	\begin{equation}
	\{(\partial \psi)^2, \Box \psi \} << V(e^{\xi}\psi)e^{-4\xi}.
	\label{slow}
	\end{equation}
	
	So, Eq. (\ref{1-6a}) can be approximated to
	\begin{equation}
	S=\frac{1}{2} \int d^{4}x \sqrt{-g} \left\{\frac{1}{k}{\mathcal{R}} -\psi^2
	g^{\alpha\beta}\partial_{\alpha} \xi \partial_{\beta}\xi-V(e^{\xi}\psi)e^{-4\xi}\right\}. \label{1-7}
	\end{equation}
	
	It is interesting to note the existence of a exponential coefficient in the potential term. If $\xi$ increases with time then the coefficient plays the role of a damping factor and  the potential decreases.

	\subsection{Power law solution in case of a monomial potential}
	As an example, we shall take a potential in the form~\cite{linde4}
	\begin{equation}
	V(\bar{\psi})= \nu m_p^4\left(\frac{\bar{\psi}}{m_p}\right)^p.
	\label{1-a8}
	\end{equation} 
	where $\nu$ and $p$ are constant quantities with $\nu<<1$. 
	
	Corresponding slow roll parameters are given by
	\begin{equation}
	\epsilon=\frac{p^2}{2\gamma^2}~,~ \eta=\frac{p(p-1)}{\gamma^2}, \label{slowroll}
	\end{equation}
	where  $\gamma \equiv \psi/m_p$ and during slow-roll inflation we can write $\gamma >> 1$, which will be discussed later in this section. 
	
	Now from Eq.~(\ref{1-7}) using the monomial potential~(\ref{1-a8}) we have
	\begin{equation}
	S=\frac{1}{2} \int d^{4}x \sqrt{-g} \left[\frac{1}{k}{\mathcal{R}} -\psi^2\left\{g^{\alpha\beta} \partial_{\alpha} \xi \partial_{\beta}\xi+ \nu m_p^{4-p}\psi^{p-2} e^{(p-4)\xi}\right\}\right]. \label{1-8}
	\end{equation}
	
	Varying the action with respect to $g_{\alpha\beta}$ and $\xi$ produces the required field equations in a spatially flat FRW background as
	\begin{equation}
	3 \left(\frac{\dot{a}}{a}\right)^2=\frac{1}{2}k
	\psi^2\left\{\dot{\xi}^2+\nu m_p^{4-p}\psi^{p-2} e^{(p-4)\xi}\right\}.
	\label{1-9}
	\end{equation}
	
	\begin{equation}
	\ddot{\xi}+3H \dot{\xi}+\frac{1}{2}(p-4)\nu m_p^{4-p}\psi^{p-2} e^{(p-4)\xi}=0.
	\label{1-10}
	\end{equation} 
	
	The corresponding solution can be provided as~\cite{kalara}
	\begin{equation}
	a(t)=a_{0} t^q \label{n},
	\end{equation}
	
	\begin{equation}
	\xi(t)=\xi_T - C\ln {\left(\frac{t}{t_T}\right)},
	\label{1-11}
	\end{equation} 
	where
	\begin{equation}
	q= 4\pi C^2\gamma^2,~C=\frac{2}{(p-4)} ~\text{and}~t_T^2=\left[\frac{4(3q-1)}{\nu \gamma^{p-2}(p-4)^2m_p^2 e^{(p-4)\xi_T}}\right].
	\label{p}
	\end{equation}
	
	In this case $\xi_T$ is a constant which is a dimensionless quantity and represents the value of $\xi$ when inflation terminates. However, following the work of Kalara~\cite{kalara} one can opt for a simplified form of Eqs.~(\ref{n}) - (\ref{p}).
	
	From Eq. (\ref{p}) we can say that the Universe during a power law inflationary phase gives $\gamma>\frac {|p-4|}{4\sqrt{\pi}}$. Here values of $\psi$ from $-\infty$ to $+\infty$ are fully appropriate. If $\rho_{\psi}$ is the energy density of $\psi$ and $\rho_{\psi}< m_p^4$, the Universe can be described classically. Since $\nu<<1$ one can constrain the kinetic energy of $\psi$, viz. $(\partial\psi)^2<m_p^4$~\cite{linde3,linde4}.
	
	From Eq. (\ref{1-11}) it is evident that $e^{(p-4)\xi}$ decreases with $t$. So $\Lambda_{eff}\equiv 4\pi\nu m_p^2\gamma^p e^{(p-4)\xi}$ reduces during inflationary phase such that $\Lambda_{eff}\sim t^{-2}$. This result matches well with the observational upper limit and also with the phenomenological models referred earlier~\cite{saibal3,saibal4} where it has been concluded that for flat Universe $\Lambda\sim t^{-2}$ is true for different models.
	
	In order to discriminate among the different components that might be responsible for the present acceleration of the universe\cite{Perlmutter,Reiss},two new geometrical parameters termed as the statefinder parameters,depending on the nature of the space-time metric were introduced by Sahni et al.\cite{Sahni}. They are usually denoted by $r$ and $s$,but here we will denote them by $r'$ and $s'$. The parameters are defined along with the deceleration parameter $q_{dec}$ as
	
	\begin{equation}
	q_{dec}=-1-\frac{\dot{H}}{H^2},
	\end{equation}  
	
	and
	
	\begin{equation}
	r'=1+\frac{3\dot{H}}{H^2}+\frac{\ddot{H}}{H^3}, s'=\frac{r'-1}{3\left(q_{dec}-\frac{1}{2}\right)}.
	\end{equation}
	
	Using the form of $a$ we obtained in Equation (31),the deceleration and statefinder parameters are obtained to be 
	
	\begin{equation}
	q_{dec}=-\left(\frac{q+1}{q}\right), r'=1+\frac{2-3q}{q^2}, s= \frac{2}{q}\left(\frac{3q-2}{3q+2}\right).
	\end{equation}
	
	We can see from the above result that despite using scalar field inflaton for generating the inflationary mechanism,both the statefinder parameters are obtained to be constants which happens to be the case for a $\Lambda$-term. Thus our description of the inflationary mechanism in terms of the $\Lambda_{eff}$ is further established.
	
	Another important point regarding the inflationary model is the exit mechanism, i.e. how inflation terminates. It has been mentioned earlier that the exit mechanism is a serious problem in PLI.  In the present work, graceful exit occurs due to the decay of vacuum density. The inflaton $\psi$ is freezed out during slow-roll paradigm and the corresponding energy density of $\psi$ is given by $\rho_{\psi}\equiv\frac{1}{2}\dot{\psi}^2+Ve^{-4\xi}\approx Ve^{-4\xi}$.  Unlike the exponential inflation, energy density $\rho_{\psi}$ in the present case does not remain constant and decays during inflation. With the evolution of time $Ve^{-4\xi}$ reduces. So at a particular stage there will be an instant of time when the kinetic term can not be neglected in $\rho_{\psi}$. At this point of time, the kinetic and the potential terms are of the same order of magnitude, i.e. the slow-roll approximation is no longer valid at this stage and the inflation terminates.
	
	After the inflationary paradigm reheating stage begins during this stage the inflaton begin oscillating near the minimum of its effective potential as a result the elementary particles are produced. These particles interact with each other and ultimately creates a state of thermal equilibrium at some temperature for the Universe. During the period of reheating the  inflaton energy is converted into matter and radiation, then the Universe re-enter the hot Big Bang model followed by dark matter and vacuum phases. In this model we deal with two dynamical scalar fields but the reheating process is actually controlled by the inflaton $\psi$. At the end of the inflation the conformal factor tends to have a constant configuration and the kinetic energy part of the inflaton, which was unimportant during inflation, becomes important. Indeed the part of $\psi$ and $\xi$ are changed during the phase of reheating and the model again  reduced to a single-field type. In the present case the reheating process proceeds in a similar manner as the standard Big Bang model does.

	\begin{figure}
		\centering
		\includegraphics[scale=0.45]{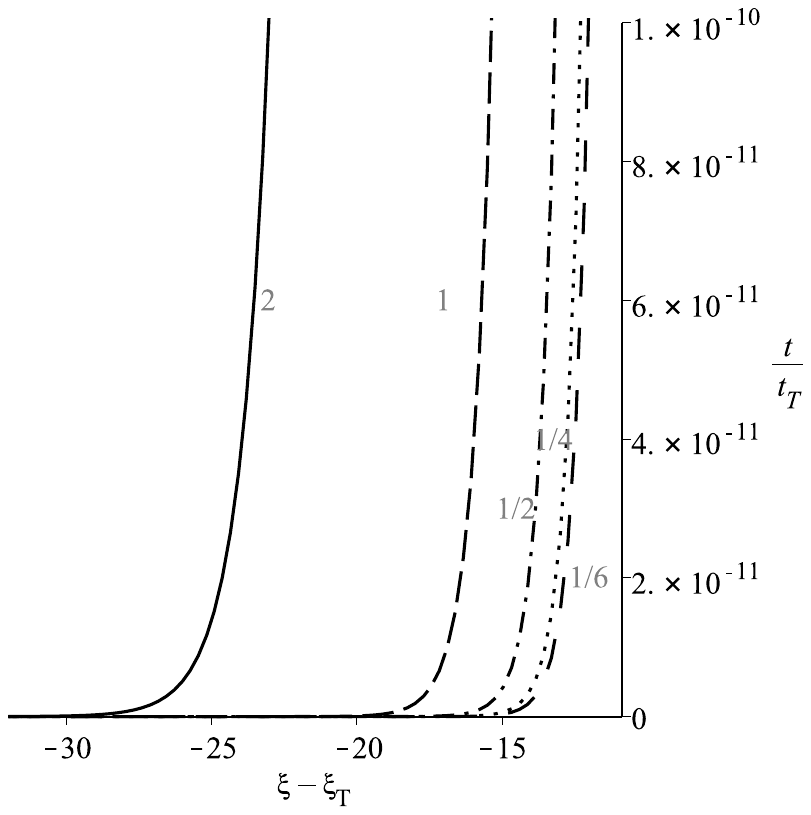}
		\vspace{-5.0cm}
		\caption{Variation of the conformal field $\xi(t)$ with time $t$ for different values of $p$.}
	\end{figure}

	Now, let us consider that inflation terminates at time $t_T$. From Eq. (\ref{1-11}) it is clear that $\xi \rightarrow \xi_0$ when $t \rightarrow  t_T$, it implies that when inflation terminates, $\xi$ takes a constant value. It is clear from Fig. 1 that practically even when $t << t_T$ variation of $\xi$ is negligible, i.e. $t_b << t_T$.  Therefore, the action given by Eq. (\ref{1-6}) reduces to
	\begin{equation}
	S=\frac{1}{2} \int d^{4}x \sqrt{-g}~ \left\{\frac{1}{k}{\mathcal{R}}-g^{\alpha\beta}
	\nabla_{\alpha} \psi \nabla_{\beta}\psi -V(\psi)\right\},
	\label{1-60}
	\end{equation}
	where the conformal factor $e^{-4\xi}$ of Eq. (\ref{1-6}) becomes a constant factor $e^{-4\xi_T}$ and hence it can be consumed by the potential. So after the termination of inflation one would expect the following two features: (i) the effective cosmological term, $\Lambda_{eff}\sim t^{-2}$ and it decreases in an identical manner like energy densities of radiation and matter dominated phase after the inflationary era in hot big bag model, and (ii) reheating initiates in a very much similar manner as the conventional inflationary models.
	
	One important point in any model regarding inflation is the the number of e-folding produced by the inflation which is defined as
	\begin{equation}
	N\equiv \ln \frac{a_T}{a_b}
	=\int_{t_b}^{t_T} \frac{da}{a}=\int_{t_b}^{t_T} H dt,
	\end{equation}
	where $t_b$ and $t_T$ are the times at which inflation begins and terminates, respectively and $a_b$ and $a_T$ are the corresponding scale factors. Using the solutions of (\ref{n}) and (\ref{1-11}) in Eq. (\ref{1-9}), we can obtain 
	\begin{equation}
	N\sim q\ln\left(\frac{t_T}{t_b}\right),
	\label{nn}
	\end{equation}
	where $q$ can be obtained from Eq. (\ref{p}). 
	
	Now to overcome the smoothness and flatness problems one must required, $N>60$.  Inspecting Eqs. (\ref{p}) and (\ref{nn}), and also the facts that $t_b << t_T$, $\gamma >> 1$, one can understand that this condition can be achieved easily.

	\begin{figure}
		\centering
		\includegraphics[scale=0.45]{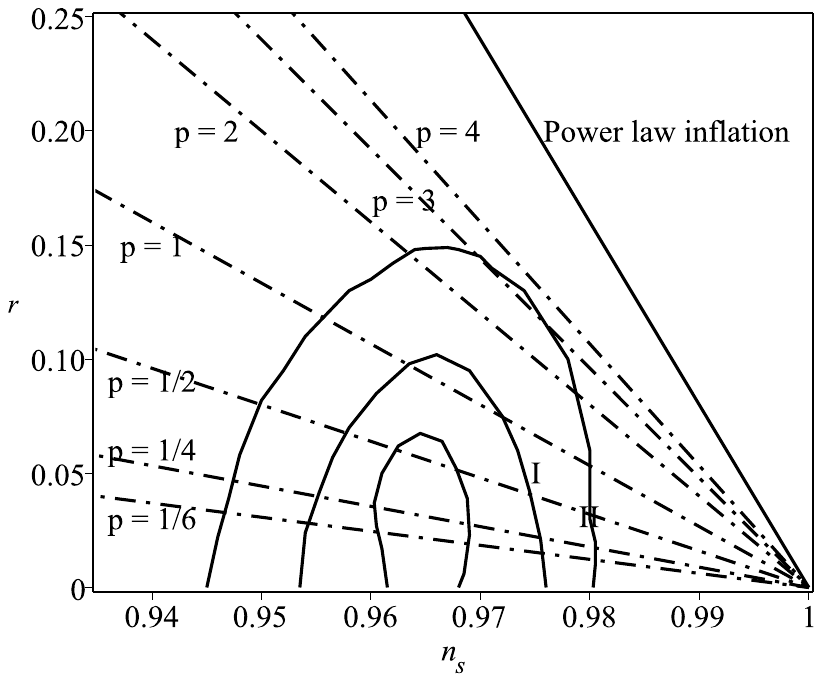}
		\vspace{-6.0cm}
		\caption{The plot of $r$ versus $n_s$ for different values of $p$ of the monomial potential (dash-dot lines). Innermost contour corresponds to the Planck + BICEP2 + Keck Array data, outermost and middle contour corresponds to the Planck + WMAP + BAO data at $\sigma$ and 2$\sigma$ confidence limits (CL) respectively. The solid line corresponds to conventional power-law inflationary models with 	exponential potential and dot lines I and II represents intermediate inflation and logamediate inflation respectively.}
	\end{figure}

	\section{Comparison with observation and other inflationary models}
	In this section we compare our model in the light of recent observational data~\cite{wmap,planck1,planck2,keck}.
	
	From Eq. (\ref{slowroll}) we have
	\begin{equation}
	n_s-1=-\frac{p(p+2)}{\gamma^2}, \label{spectral} 
	\end{equation}
	
	\begin{equation}
	r =\frac{8p^2}{\gamma^2}.\label{scavec}
	\end{equation} 
	
	Equations (\ref{spectral}) and (\ref{scavec}) are plotted in Fig. 2 for different values of $p$. 
	
	Fig. 2 shows that while conventional power law inflationary model with exponential potential remains completely outside the region allowed by observational results in $\{r, n_s\}$ space, the present model (\ref{1-8}) is in very good agreement with~\cite{wmap,planck1} and also fairly well with~\cite{planck2,keck} along with these highly modified inflationary models. It is worth noting that in the present work for minimal coupling scalar field power-law potential index must satisfy $p<4$ in order to satisfy the observational data~\cite{wmap,planck1,planck2,keck}. It is different from non-minimal coupling scalar field case for which power law potential index must satisfy $p>4$~\cite{bairagi}. In general, for chaotic inflationary models usual convention is that $p \geq 1$, however in the present model one can see that for $p<1$ latest experimental result can be retrieved very well. In this connection it is worthwhile to mention that the monomial potentials $V(\psi) = \nu m^4_{Pl} (\psi/m_{Pl})^p$ as provided [vide Eq. (\ref{1-a8})] by Linde~\cite{linde4} with $p \geq 2$ are strongly unfavorable with respect to the $R^2$ model. It is argued~\cite{planck2} that for these values the Bayesian evidence is worse than in 2015 because of the smaller level of tensor modes allowed by BK14~\cite{Ade2018}. Models with $p = 1$ or $p = 2/3$ are more compatible with the data~\cite{SW2008,McAllister2010,McAllister2014}. It is interesting to note that our prescribed value $p<1$ is well within the second option $p = 2/3$.

	\section{Conclusion}
	Conventional PLI with exponential potential has important limitations: cosmological constant problem, graceful exit problem and the mismatch of the parameters obtained theoretically with recent observational results. Over the years different inflationary models have been proposed to solve some cosmological problems. In the standard inflationary models people use a minimally coupled canonical scalar field having an appropriate potential function. Some of these inflationary models depends on the existence of a large cosmological constant at early times but they do not predict anything about the its smallness at late times. 
	
	In our model natural explanations to these drawbacks can be obtained. We have probed a single scalar field, termed as inflaton whose potential can receive contributions of masses from different fields in Standard Model. Thus it gives a large value of the effective cosmological constant. Here, two metrics in the gravitational and the inflaton parts pertain to two different conformal frames. It has been shown that this type of anomalous gravitational coupling of inflaton has novel features in inflationary paradigm.  We can jot down the main results as follows :\\
	1) The conformal factor behaves in this model like a dynamical field and the anomalous coupling provides a damping factor which is bestowed to the effective potential. Due to the decay mechanism the effective cosmological term reduces with the time evolution and hence the large cosmological constant decreases during inflationary phase. Thus the cosmological constant problem in case of PLI is pacified.
	
	2) A set of exact solutions in case of a monomial inflaton potential are obtained, resulting power law inflation. From the solution we can conclude that inflation terminates at a particular time and with the evolution of time the radiation and matter dominated era after inflation in the standard hot big bang model occur. Thus graceful exit problem can be eased. 
	
	3)We can effectively describe the inflationary mechanism in terms of $\Lambda_{eff}\equiv 4\pi\nu m_p^2\gamma^p e^{(p-4)\xi}$. The statefinder parameters $r'$ and $s'$ turn out to be constants for the scale factor obtained by us from power-law inflation with a monomial potential,as should be the case for acceleration due to a $\Lambda$- term.  
	
	4) We have shown that the values of the spectral index and the tensor-to-scalar ratio are in very good agreement with the observational data. Comparison has been made with a non-minimally coupled to scalar field situation.

	\section*{Acknowledgments} SR is thankful to the Inter-University Centre for Astronomy and Astrophysics (IUCAA), Pune, India for providing Visiting Associateship under which a part of this work was carried out. PP is specially grateful to Prof. Amit Kundu, IIEST for encouraging discussion.

\end{document}